\documentstyle[prl,aps,multicol,psfig]{revtex}

\begin{document}

\draft

\title{Localization in Artificial Disorder - Two Coupled Quantum Dots}

\author{M. Brodsky\footnote{Email: \it misha@electron.mit.edu}, N.B.
Zhitenev\footnote{Present address: \it Bell Laboratories, Lucent
Technologies, Murray Hill, NJ 07974 }, R.C. Ashoori}

\address{Department of Physics, Massachusetts Institute of Technology,
Cambridge, MA 02139}

\author{L.N. Pfeiffer, K.W. West}

\address{Bell Laboratories, Lucent Technologies, Murray Hill, NJ
07974}

\maketitle

\begin{abstract}

Using Single Electron Capacitance Spectroscopy, we study electron
additions in quantum dots containing two potential minima separated by
a shallow barrier. Analysis of addition spectra in magnetic field
allows us to distinguish whether electrons are localized in either
potential minimum or delocalized over the entire dot. We demonstrate
that high magnetic field abruptly splits up a low-density droplet into
two smaller fragments, each residing in a potential minimum. An
unexplained cancellation of electron repulsion between electrons in
these fragments gives rise to paired electron additions.

\end{abstract}


\begin{multicols}{2}

For half a century physicists have worked to understand localization
of strongly interacting electrons in a disorder potential. Either
electron interactions or disorder can produce
localization\cite{Mott,Anderson}. Though their interplay in
two-dimensional systems has been a subject of intense experimental and
theoretical studies\cite{MIex,MIth}, no theory exists fully describing
the effects of both disorder and strong interaction.

Quantum dots provide a convenient system for studying electron
localization on a microscopic scale. However, the traditional
transport techniques for studying lateral quantum dots\cite{MK-R}
sense primarily delocalized electronic states. A possible exception is
transport studies in vertical structures, but these do not permit
variation of electron density\cite{T-SH}, a critically important
parameter that changes the effective strength of electron
interactions. We study electron additions in vertical quantum dots
using Single Electron Capacitance Spectroscopy (SECS)\cite{R-L}. It
has demonstrated the capability of probing both {\it localized} and
delocalized states of electrons. Furthermore, this method allows us to
study 2D dots of various sizes and over a broad range of electron
densities.

In quantum dot experiments in high-density dots, the Coulomb repulsion
between electrons largely sets the amount of energy required to add an
additional electron to the dot. This energy increases by a fixed
amount with each electron added. An external gate, capacitively
coupled to the dot, can then be used to change the electron number,
and electron additions occur periodically in the gate voltage with a
period $e/C_{g}$, where $C_{g}$ is the capacitance between the gate
and the dot\cite{MK-R}.

In contrast, our prior SECS measurements have shown that the {\it
low-density} regime appears entirely different. The addition spectrum
of a 2D-electron droplet larger than $0.2 \mu m$ in diameter and below
a critical electron density ($n_{0} = 1 \times 10^{11} cm^{-2}$ in all
of our samples) is highly nonperiodic. It contains pairs and bunches:
two or more successive electrons can enter the dot with nearly the
same energy\cite{R-L,NZ-B}. The paired electrons thus show almost no
sign of repelling each other. Application of high perpendicular
magnetic field increases $n_{0}$ linearly, creating a sharp {\it
boundary} between periodic and ``paired" parts of the addition
spectrum\cite{NZ-B}. We hypothesized that, for densities below this
boundary, disorder and electron-electron interactions within the
low-density droplet split it into two or more spatially separate
droplets, and pairing arises once this localization occurs. We have
produced experiments to study this localization-delocalization
transition in a controlled fashion. One recently established the
existence of electronic states localized at the dot's periphery and
arising at densities just below the critical density
$n_{0}$\cite{NZ-SG}.

In this letter we report the results of a new approach for studying
localization and pairing in quantum dots. We intentionally create a
dot with an artificial ``disorder" potential: a potential profile
containing two smooth minima separated by a barrier, as in the double
dot system described below. Through analysis of addition spectra in
magnetic field, we distinguish between electrons localized in either
potential well or delocalized over the entire dot. Our studies
conclusively demonstrate that under precisely the same conditions for
observation of the paired electron additions, a low-density electron
droplet inside the dot indeed splits up into smaller fragments. This
abrupt disintegration creates a sharp {\it boundary} between periodic
and ``paired" parts of the addition spectra, with paired electrons
entering into spatially distinct regions within a dot. We also measure
the remnant residual interaction between the fragments. Surprisingly,
it displays a nearly complete independence on the strength of the
applied field for fields larger than required for the localization
transition. While no theory exists explaining the observed transition
or the pairing phenomenon, recent numerical simulations display
results similar to some of our data\cite{Canali}.

The dots were fabricated within an AlGaAs/GaAs heterostructure as
described in previous work\cite{R-L,NZ-B}. The essential layers (from
bottom to top) are a conducting layer of GaAs serving as the only
contact to the dots, a shallow AlGaAs tunnel barrier, a GaAs active
layer that contains the dots, and an AlGaAs blocking layer. On the top
surface, we produce a small AuCr top gate using electron beam
lithography. This top gate was used as a mask for reactive ion etching
that completely depletes the active GaAs layer in the regions away
from the AuCr gate.

To create a barrier within a dot we pattern a top gate in a dumbbell
shape. This produces two small vertical dots laterally separated by a
small distance (a schematic of our samples is shown in the Fig.1B).
The top gate controls the electron density of the entire system. This
geometry results in a double potential well with two valleys separated
by a saddle. By changing the top gate bias $V_{g}$, we gradually fill
the double dot system with electrons. At first electrons accumulate in
two independent electron puddles, one localized in each dot. The
puddles grow laterally with increasing electron number and eventually
couple to each other. The coupling mixes states of one dot with those
of the other, and electrons start traversing the saddle point. When
the two puddles finally merge into a single large dot, the electron
wave functions spread over the entire area of the resulting large dot.

By varying lithographic dimensions, we control the height of the
saddle and therefore the individual dot electron density at which
merging occurs. We examine a number of samples to investigate a broad
range of such densities: from two dots each containing a few localized
electrons up to densities $n = 2.5 - 3.5 \times 10^{11} cm^{-2}$ in
each dot.

The measurements are carried out using on-chip bridge circuit
described in\cite{R-L}. To register electron additions, we monitor the
a.c. capacitive response to a small ($ < 80 \mu V$) a.c. excitation
applied between the top gate and the contact layer. Since one top gate
covers both individual dots, an electron addition to either of the
dots results in a peak\cite{R-L} in our capacitance measurements.

To distinguish electrons added to one dot from those added to the
other, we follow the evolution of the addition spectrum with
perpendicular magnetic field. The general behavior of the electron
addition spectrum for a single dot in magnetic field is well known
both for case of few-electron droplets\cite{R-S,T-SH} and for
many-electron dots in Quantum Hall regime\cite{McE,OK}. Addition
energies oscillate with field as electrons shift between different
angular momentum states. The exact pattern of those oscillations
depends sensitively on the details of the confinement potential, and
serves as a ``signature" of a particular dot. Although in our samples
the two dots are made to be nominally identical, the particular shapes
of the confinement potential of the two dots are slightly different
due to disorder and imperfections in the lithography process. Addition
energies for the two dots thus depend differently on the
perpendicularly applied magnetic field, permitting us to associate
each electron addition with a particular dot.

The capacitance traces taken at different values of the magnetic field
are plotted together on the greyscale panel in Fig.1A. Black denotes
high capacitance. Each successive trace corresponds to the energy for
adding an electron to the double dot system. The lowest trace shown
represents the first electron added to the two-dot system. The
low-density part of the spectrum ($-290 mV < V_{g} < -135 mV$) appears
as a simple superposition of two different families of traces. First
10 electron addition traces comprising one family are marked by
dashes. Each family can be described qualitatively within the constant
interaction model for Darwin-Fock states, as is typical for individual
small circular dots\cite{T-SH,LK-R,R-S}. Because such separation of
the spectrum is possible, we conclude that up to $V_{g}=-135 mV$ our
system consists of two independent electron droplets. Incidental
alignment of the ground states of the two droplets for some particular
values of the gate bias and the magnetic field may cause simultaneous
but independent electron additions to each individual dot. Indeed,
multiple level crossings (some marked by circles on Fig.1A) can be
seen on the plot. At each crossing point the peak in the capacitance
signal has double height, indicating an independent addition of two
electrons to the two-dot system. The exact coincidence of the peaks
suggests that capacitive coupling between two droplets is negligible.
At much higher densities ($V_{g}>-45 mV$) there is only one periodic
Coulomb ladder, indicating that the initially separate electron
droplets have merged into a single one.

The transition between the two limits occurs over gate biases $-135
mV<V_{g}<-45 mV$, depending on the strength of the applied magnetic
field. At zero field, the merging occurs in an interval $ \Delta
V_{g}=25 mV$ wide centered around $V_{g}=-125 mV$. The gate bias
$V_{g}=-125 mV$ corresponds to electron densities in each individual
dot of $1.2 \times 10^{11}cm^{-2}$ and $1.7 \times 10^{11}cm^{-2}$
respectively. Each dot contains about 30 electrons. For higher
densities and at zero field there is one combined dot under the gate.
However, magnetic field greater than $4T$ dramatically affects the
spectrum. There exist a clearly visible sharp boundary, which
separates the spectrum in two parts. It is marked by a line on Fig.1A.
To the left of the boundary (the low field side), all electron
addition traces show similar evolution with magnetic field; electrons
appear to enter one combined dot and Coulomb blockade produces nearly
periodic addition spectrum. To the right of the boundary (the high
field side), the addition traces are grouped into bunches. With
increasing magnetic field, the boundary between the two regimes
extends up to densities of $1.7 \times 10^{11}cm^{-2}$ and $2.2 \times
10^{11}cm^{-2}$, in each dot respectively (over 60 total electron
additions to the two-dot system). An increase in density of each dot
along the boundary can be approximated by the linear relation $ \Delta
n \propto 0.1 \times B(T) \times 10^{11}cm^{-2}$ for both of the two
individual dots.  This linear relation holds for all of our samples.
Surprisingly this boundary follows the same linear density-field
relation as the one seen in individual dots of larger
sizes\cite{NZ-B}.

To understand the origin of this boundary we expand the addition
spectrum to the right of the boundary (Fig.1C) and focus on six marked
subsequent addition traces $R1$, $B1$, $H1$, $H2$, $B3$, $R3$. All of
the marked traces again oscillate with magnetic field. But here the
origin of the oscillations is different from that of the few electron
case considered above. For magnetic field higher than $4T$, electrons
within each dot fill only the lowest orbital Landau level, but with
both spin-up and spin-down electrons. With increasing magnetic field,
the electron orbits shrink and Coulomb repulsion causes redistribution
of electrons between the two spin-split branches of the lowest orbital
Landau level. This produces oscillations in the single electron traces
known as ``spin flips"\cite{McE,OK,MK-R}.

Fig.1C shows two different oscillation patterns. One is represented by
traces $R1$ and $R3$; similarly, traces $B1$ and $B3$ display another
pattern. The existence of two patterns characteristic of the
individual dots indicates that to the right of the boundary there
exist two separate dots, despite the fact that for zero field two dots
are merged into one. We conclude that the boundary separates two
regimes in $V_{g}-B$ space. In one regime, electron wavefunctions are
spread over the entire area of the double dot and in the other each
electron dwells in one of two individual dots.

In the latter regime, the two dots are not completely independent.
Though magnetic field breaks one combined electron dot into two
separate ones, residual coupling remains. The barrier between the two
dots is small, and interdot tunneling remains possible\cite{CL-S}.
When ground states of individual dots are aligned with each other a
finite tunnel coupling splits two aligned levels\cite{MG,GH}. Such
alignment creates the equivalent of a molecular hybrid state, which
appears as a bunch in the spectrum. An example of such splitting are
the two hybridized traces in the middle of the plot on the Fig.1C:
$H1$, $H2$. They cannot be solely associated with either of the two
spin-flip patterns but rather exhibit features belonging to both of
them.

\begin{figure} 
\narrowtext 
\psfig{figure=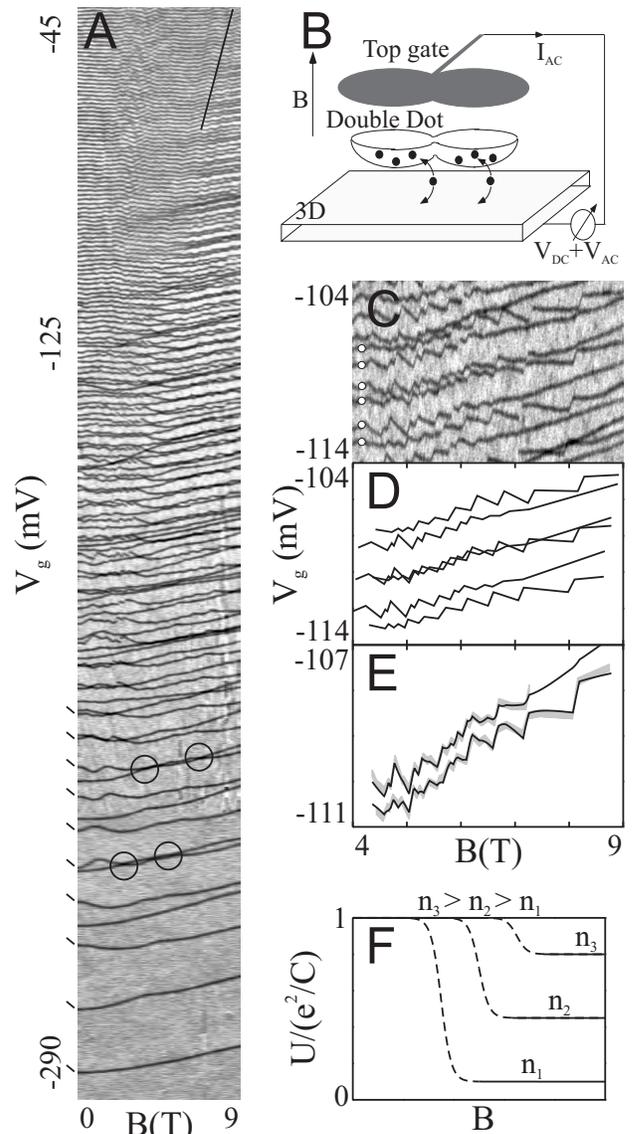,height=5.97in} 
\caption{\narrowtext 
{\bf A.} Schematic of our samples. The dots
potential profile contains two minima separated by a barrier. A single
top gate controls the electron densities of the entire two-dot system.
{\bf B.} Greyscale plot of quantum dot capacitance as a function of
gate bias and magnetic field. Black denotes high capacitance. Each
successive trace corresponds to the energy for adding an electron to
the double dot system. First 10 additions to one dot are marked by
dashes. Circles mark level crossings. A diagonal line delineates a
clearly visible sharp {\it boundary} described in the text. {\bf C.}
An addition spectrum expanded to the right of the boundary. Six
subsequent addition traces marked by empty circles are $R1$, $B1$,
$H1$, $H2$, $B3$, $R3$ (bottom to top). $R1$, $R3$ and $B1$, $B3$
represent two oscillation patterns. Hybridized traces $H1$, $H2$ do
not belong to any of the patterns. {\bf D.} The hypothetical spectrum
in absence of the interaction between to dots. Hybrid states $H1$,
$H2$ are replaced by two unperturbed independent states from two
dots:$R2$, $B2$. {\bf E.} Reconstruction of the hybrid states. The
data ($H1$, $H2$) are shown in grey, black are fits. {\bf F.}
Schematic field dependence of the tunneling matrix element $U$ for
different densities ($n_{1}<n_{2}<n_{3}$). Solid lines denote regions
where residual coupling is extracted as described in the text.}
\end{figure}

We estimate the coupling strength between two dots by describing the
spectra using single particle states. We reconstruct the two
hybridized states $H1$ and $H2$ from the neighboring ``one-dot states"
$R1$, $B1$, $B3$, $R3$ a following way. First, we assume that in the
absence of the residual interaction the spectrum would be as presented
in Fig.1D. In place of the hybrid states $H1$, $H2$ there are two
unperturbed independent states from the two dots: $R2$ and $B2$ . For
these unperturbed states we take $R2=(R1+R3)/2$ and $B2=(B1+B3)/2$ .
Tunneling between $R2$ and $B2$ produces an off-diagonal matrix
element $U$. Diagonalization of the Hamiltonian 
$\left[\begin{array}{cc} B2 & U \\ U^{*} & R2 \end{array} \right] $ 
splits $R2$ and $B2$ into: 
\begin{equation} 
E=\frac{(R2+B2)}{2} \pm \sqrt{\frac{(R2-B2)^{2}}{4}+U^{2}} 
\end{equation}

Surprisingly for such a simplistic model using $U$ as the only fitting
parameter, we obtain practically perfect fits to our data (Fig.1E).
Unexpectedly, the residual coupling strength $U$ ($U \approx 0.1
\times (e/C_{dot}$) for the case shown) displays nearly complete
independence of the strength of the applied field for fields larger
than required for the localization transition.

The results of similar fitting for different densities are summarized
in Fig.1F. Though constant in field, this coupling increases with
density, and becomes comparable to $E_{c}=e/C_{dot}$ at densities
around $2 \times 10^{11}cm^{-2}$. The boundary ceases to exist at
these densities. In fact, the boundary is altogether absent in samples
for which the individual dot densities at the merging point are higher
than $2.3 \times 10^{11}cm^{-2}$, i.e. magnetic field has no effect on
the merging of two high-density dots.

Our data convincingly establish that high magnetic field abruptly
splits a low-density electron droplet placed in disorder potential
into smaller fragments. It is this split up that causes a sharp
boundary in the addition spectrum. The paired electron additions to
the dot seen to the right of the boundary result from an unexplained
cancellation of electron repulsion between electrons in these
fragments. The boundary essentially separates two phases: in one,
electrons are delocalized over entire sample, and in the other,
electrons are confined in local disorder minima.

The physical mechanism of such separation or of the pairing phenomena
has yet to be established. However, recent preprint\cite{Droplet}
shows that a two-phase coexistence of high density liquid and a
low-density gas might be energetically favorable in the interacting
two-dimensional system placed in disorder potential, and numerical
calculations by Canali\cite{Canali} support our finding that two
electrons in the pair enter into spatially separated regions of the
dot.

We would like to acknowledge useful discussions with C. de C. Chamon,
G. Finkelstein, D. Goldhaber-Gordon, B.I. Halperin, M.A. Kastner, L.S.
Levitov and K.A. Matveev. Expert etching of samples was performed by
S.J.Pearton. This work is supported by the ONR, JSEP-DAAH04-95-1-0038,
the Packard Foundation, NSF DMR-9357226 and DMR-9311825.

\end{multicols}



\begin{references}

\bibitem{Mott} N.F. Mott, Proc.Phys.Soc.London, Ser. A 62, 416 (1949).

\bibitem{Anderson} P.W. Anderson, Phys.Rev. {\bf 109}, 1948 (1958).

\bibitem{MIex} S.V. Kravchenko, G.V. Kravchenko, J.E. Furneaux, V.M.
Pudalov, M. D'Iorio, Phys.Rev.B {\bf 50}, 8039, (1994); D. Popovic,
A.B. Fowler and S. Washburn, Phys.Rev.Lett. {\bf 79}, 1543 (1997).

\bibitem{MIth} D. Belitz and T.R. Kirkpatrick, Rev.Mod.Phys. {\bf 66},
261 (1994); B.L. Altshuler and A.G. Aronov, in {\it Electron-Electron
Interaction in Disordered Systems}, edited by M. Pollak, A.L. Efros,
(North Holland, Amsterdam 1985); P.A. Lee, T.B. Ramakrishnan,
Rev.Mod.Phys. {\bf57}, 287 (1985).

\bibitem{R-L} R.C. Ashoori, H.L. Stormer, J.S. Weiner, L.N. Pfeiffer,
S.J. Pearton, K.W. Baldwin, K.W. West, Phys.Rev.Lett. {\bf 68}, 3088
(1992); R.C.Ashoori {\it et al.}, Physica B, {\bf 189}, 117 (1993).

\bibitem{MK-R} M.A.Kastner, Physics Today {\bf 46}, 24 (1993).

\bibitem{R-S} R.C. Ashoori, H.L. Stormer, J.S. Weiner, L.N. Pfeiffer,
K.W. Baldwin, K.W. West , Phys.Rev.Lett. {\bf 71}, 613 (1993);
R.C.Ashoori, Nature {\bf 379}, 413 (1996).

\bibitem{NZ-B} N.B. Zhitenev, R.C. Ashoori, L.N. Pfeiffer, K.W. West,
Phys.Rev.Lett. {\bf 79}, 2308 (1997); R.C. Ashoori {\it et al.},
Physica E {\bf 3}, 15 (1998).

\bibitem{NZ-SG} N.B. Zhitenev, M. Brodsky, R.C. Ashoori, L.N.
Pfeiffer, K.W. West, Science {\bf 285} 715 (1999).

\bibitem{LK-R} For a review see L.P. Kouwenhoven, C.M Marcus, P.L.
McEuen, S. Tarucha, R.M. Westervelt, N.S. Windgreen in {\it Mesoscopic
Electron Transport}, edited by L.L. Sohn, L.P. Kouwenhoven, G.
Sch\"on, NATO ASI, Ser. E, (Kluwer Academic Publishers, Dordrecht,
1997); and references therein.

\bibitem{T-SH} S. Tarucha, D.G. Austing, T. Honda, R.J. van der Hage,
L.P. Kouwenhoven, Phys.Rev.Lett. {\bf 77}, 3613 (1996).

\bibitem{McE} P.L. McEuen, E.B. Foxman, J. Kinaret, U.Meirav, M.A.
Kastner, N.S. Wingreen, S.J. Wind Phys.Rev.B {\bf 45}, 11419 (1992).

\bibitem{OK} O. Klein, C. de C. Chamon, D. Tang, D.M. Abusch-Magder,
U. Meirav, X.-G. Wen, M.A. Kastner, S.J. Wind Phys.Rev.Lett. {\bf 74},
785 (1995).

\bibitem{CL-S} C. Livermore, C.H. Crouch, R.M. Westervelt, K.L.
Campman, A.C. Gossard, Science, {\bf 274}, 1332 (1996) and references
therein.

\bibitem{MG} K.A. Matveev, L.I. Glazman, H.U. Baranger, Phys.Rev. B
{\bf 53}, 1034 (1996); Phys.Rev.B {\bf 54}, 5637 (1996).

\bibitem{GH} J.M. Golden and B.I. Halperin, Phys.Rev.B {\bf 53}, 3893
(1996); Phys.Rev.B {\bf 54}, 16757 (1996).

\bibitem{Droplet} Junren Shi, Song He and X.C. Xie, cond-mat/9909450.

\bibitem{Canali} C.M. Canali, cond-mat/9909220.

\end{references}
\end{document}